\documentclass[10pt,conference]{IEEEtran}

\IEEEoverridecommandlockouts
\usepackage{cite}
\usepackage{amsmath,amssymb,amsfonts}
\usepackage{algorithmic}
\usepackage{graphicx}
\usepackage{textcomp}
\usepackage{xcolor}
\usepackage{booktabs}
\usepackage{hyperref}
\usepackage{listings}
\usepackage[bottom]{footmisc}
\usepackage{xspace}
\usepackage{xurl}
\usepackage{tabularx}
\usepackage{multirow}
\usepackage{todonotes}
\usepackage[acronym]{glossaries}
\usepackage[english]{babel}
\usepackage[autolanguage]{numprint}
\usepackage{endnotes}
\usepackage{pbox}

\def\Csharp{C\raisebox{0.5ex}{\tiny\textbf{\#}}\xspace}

\newcommand{\citeurl}[1]{\endnote{{\scriptsize \url{#1}}}}

\newcommand{\toolname}[1]{{\scriptsize \texttt{#1}}}
\newcommand{\license}[1]{\emph{#1}}

\newcommand{\gpltwo}{\license{GPL-2.0}\xspace}
\newcommand{\gpl}{\license{GPL}\xspace}
\newcommand{\mymit}{\license{MIT}\xspace}
\newcommand{\agpl}{\license{AGPL}\xspace}
\newcommand{\apache}{\license{Apache}\xspace}

\newcommand{\cargo}{\toolname{Cargo}\xspace}
\newcommand{\maven}{\toolname{Maven}\xspace}
\newcommand{\pypi}{\toolname{PyPI}\xspace}
\newcommand{\ruby}{\toolname{RubyGems}\xspace}
\newcommand{\npm}{\toolname{NPM}\xspace}
\newcommand{\nuget}{\toolname{NuGet}\xspace}
\newcommand{\packagist}{\toolname{Packagist}\xspace}

\newcommand{\pr}{\emph{PageRank}\xspace}

\newacronym{npm}{NPM}{Node Package Manager}
\newacronym{pypi}{PyPI}{Python Package Index}
\newacronym{rails}{\toolname{rails}}{Ruby on Rails}
\newacronym{gpl}{\license{GPL}}{GNU General Public License}
\newacronym{agpl}{\license{AGPL}}{GNU Affero General Public License}
\newacronym{spdx}{SPDX}{Software Package Data Exchange}
\newacronym{pr}{PR}{Plub}
\newacronym{cs}{CS}{Plab}
\newacronym{tf}{TF}{Plob}

\def\BibTeX{{\rm B\kern-.05em{\sc i\kern-.025em b}\kern-.08em
    T\kern-.1667em\lower.7ex\hbox{E}\kern-.125emX}}
\begin{document}
\renewcommand{\tableautorefname}{Tab.}
\renewcommand{\figureautorefname}{Fig.}
\renewcommand{\sectionautorefname}{Sec.}
\renewcommand{\subsectionautorefname}{Sec.}
\renewcommand{\subsubsectionautorefname}{Sec.}
\renewcommand{\equationautorefname}{Eq.}

\title{License Incompatibilities in Software Ecosystems}

\author{\IEEEauthorblockN{Rolf-Helge Pfeiffer}
\IEEEauthorblockA{ %
\textit{IT University of Copenhagen}\\
Copenhagen, Denmark \\
ropf@itu.dk}
}

\maketitle

\begin{abstract}
Contemporary software is characterized by reuse of components that are declared as dependencies and that are received from package managers/registries, such as, \npm, \pypi, \ruby, \toolname{Maven Central}, etc.
Direct and indirect dependency relations often form opaque dependency networks, that sometimes lead to conflicting software licenses within these.

In this paper, we study license use and license incompatibilities between all components from seven package registries (\cargo, \maven, \npm, \nuget, \packagist, \pypi, \ruby) with a closer investigation of license incompatibilities caused by the \gls{agpl}.

We find that the relative amount of used licenses vary between ecosystems (permissive licenses such as \mymit and \apache are most frequent), that the number of direct license incompatibilities ranges from low 2.3\% in \cargo to a large 20.8\% in \pypi, that only a low amount of direct license incompatibilities are caused by \agpl licenses (max. 0.04\% in \pypi), but that a whopping 6.62\% of \maven packages are violating the \agpl license of an indirect dependency.
Our results suggest that it is not too unlikely that applications that are reusing packages from \pypi or \maven are confronted with license incompatibilities that could mean that applications would have to be open-sourced on distribution (\pypi) or as soon as they are publicly available as web-applications (\maven).
\end{abstract}

\section{Introduction}\label{sec:intro}

Denmark is one of the most digitized countries in Europe~\cite{desi2019}.
After implementation of the 2011 Danish e-government strategy~\cite{danmark2011digital}, citizens interact with the government mostly through e-government systems, such as, self-service web-applications and websites.
Examples for such systems are \toolname{Sundhed.dk}\citeurl{https://www.sundhed.dk/} (health.dk) or \toolname{Borger.dk}\citeurl{https://www.borger.dk} (citizen.dk).
These e-government systems are usually implemented as closed-source systems by private contractors and often in a combination of \Csharp, Java, JavaScript, etc.

Besides application of modern programming languages, contemporary software development relies to a large degree on \emph{code reuse}.
Software is usually build with the help of reusable components, which we call \emph{packages} in this paper.
Programming language ecosystem emerge usually around specific package managers, which download and setup needed dependencies from remote package registries, such as, \toolname{NPM}, \toolname{Maven Central}, \toolname{PyPI}, \toolname{RubyGems}, etc.

Code reuse via dependency on external packages is also encouraged by software quality models and tools, such as the SIG/TÜViT Evaluation Criteria for Trusted Product Maintainability~\cite{sig10criteria}.
It recommends using \emph{``libraries and frameworks over 'homegrown' implementations of standard functionality''}~\cite{visser2016building} for example to keep the size of software under own maintenance small.

However the advantages of code reuse, e.g., increased software quality or developer productivity~\cite{basili1996reuse,lim1994effects,mohagheghi2004empirical} are attended by drawbacks, such as, increased maintenance cost or legal issues
~\cite{orsila2008update,vendome2018assisting,mathur2012empirical,duan2017identifying}.

This paper focuses on license incompatibilities, which are examples of such legal issues.
For instance, in March 2020 it was reported that \numprint{577148} software projects which are build on top of \gls{rails} face license issues\citeurl{https://www.theregister.com/2021/03/25/ruby_rails_code/}.
An indirect dependency of the web-application framework \toolname{rails} (see \toolname{mimemagic} in \autoref{fig:deps}) switched its license from the permissive \license{MIT} to the more restrictive \gpltwo.
The restrictive strong copyleft \gls{gpl} is viral in that all work that is distributed and that includes code under \gls{gpl} has to adopt the same --or a compatible-- license.
Consequently, all dependents of \toolname{mimemagic}, i.e., \toolname{rails} and all applications build with it, would have to either switch to \gls{gpl} too or they would all have to replace the indirect dependency \toolname{mimemagic}, with a suitably licensed package.
Such a small \gpl licensed package hidden in a dependency network would pose a legal issue for commercial vendors that create and ship software on top of \toolname{rails}.
Usually they are not interested in open-sourcing their products on distribution, which is a requirement of the \gls{gpl}.
The younger \gls{agpl}\citeurl{https://tldrlegal.com/license/gnu-affero-general-public-license-v3-(agpl-3.0)} is even more restrictive on when to open-source software that reuses correspondingly licensed packages.
It prescribes that software has to be made publicly available when it accessed via a network and when it relies on \gls{agpl} licensed components.
Consequently, if one of Denmark's online e-government systems had a direct or indirect dependency to an \gls{agpl} licensed package, either it would have to be open-sourced or the respective dependency would have to be replaced with software under a compatible license.

To understand to which degree license incompatibilities pose an issue in seven major software ecosystems (\cargo (Rust), \maven (Java and other JVM languages), \npm (mostly JavaScript), \nuget (\Csharp and other .Net languages), \packagist (PHP), \pypi (Python), \ruby (Ruby)), we investigate empirically and quantitatively the following research questions on the \emph{Libraries.io} dataset~\cite{katz2020_3626071}.

\begin{description}
    \item[RQ1] Which licenses are mainly used in various software ecosystems?
    \item[RQ2] How many direct license incompatibilities exist in various ecosystems?
    \item[RQ3] How many direct license incompatibilities are caused by \gls{agpl} licenses?
    \item[RQ4] What is the impact of indirect \gls{agpl} license  incompatibilities?
\end{description}

Our goal is to investigate how many cases similar to the motivational \toolname{rails} example exist in various ecosystems and to assess if there is a risk for danish online e-government systems.

Our results show that permissive licenses, such as, \mymit and \apache are most frequent in the studied ecosystems.
The number of direct license incompatibilities is not negligible, ranging from low 2.3\% in \cargo over 4.7\%
, 5.2\%, 6.2\%, 9.1\%, and 9.1\% in \nuget, \npm, \ruby, \maven, and \packagist respectively to a large 20.8\% in \pypi.
Only a low amount of direct license incompatibilities are caused by \agpl licensed packages (max. 0.04\% in \pypi).
Lastly, a whopping 6.62\% of \maven packages are violating the \agpl license of an indirect dependency.
However, only a few packages are responsible for this large number of indirect license incompatibilities in \maven.

In the remainder, we describe the impact of license incompatibilities and potential resolutions more detailed in \autoref{sec:background}, we detail the our experiment setup in \autoref{sec:method}, we describe our results in \autoref{sec:results}, and we discuss our results in \autoref{sec:discussion} and \autoref{sec:conclusions}.

\section{Background}\label{sec:background}

\begin{figure}[b]
    \vspace{-1em}
    \centering
    \vspace{0.1em}
    \includegraphics[width=\linewidth]{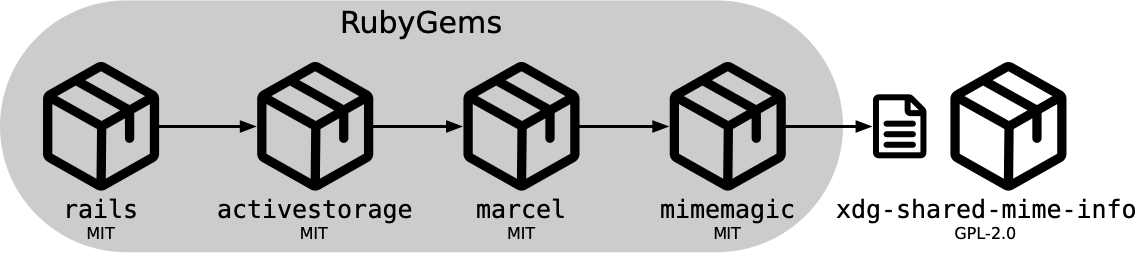}
    \vspace{-1em}
    \caption{Direct and indirect dependencies of \gls{rails} with a license incompatibility between \mymit and \license{GPL-2.0}.}
    \label{fig:deps}
\end{figure}

\textbf{License Incompatibility Example:}
\autoref{fig:deps} illustrates an excerpt of the dependencies of the web-framework Ruby on Rails (\toolname{rails}) before Mar. 24th 2021.
Via the package manager \toolname{RubyGems}, \toolname{rails} depends indirectly (by \toolname{activestorage} and \toolname{marcel}) on \toolname{mimemagic}, a library for detecting MIME types of files.
\toolname{mimemagic} in turn depends on a single artifact from the \toolname{xdg-shared-mime-info}\citeurl{https://github.com/freedesktop/xdg-shared-mime-info} project.
That artifact, is an XML file\citeurl{https://github.com/freedesktop/xdg-shared-mime-info/blob/master/data/freedesktop.org.xml.in} that serves as database of MIME types, which is copied from the original project and reused in \toolname{mimemagic}.
The XML file and the entire \toolname{xdg-shared-mime-info} are licensed under \emph{GPL-2.0} whereas \toolname{mimemagic} is under \emph{MIT} license.
These two licenses are incompatible, since \emph{GPL-2.0} is a so-called \emph{"viral"} license that would require dependents to open-source their code under the same license, which the \emph{MIT} license does not require.

After being informed about the license incompatibility\citeurl{https://github.com/mimemagicrb/mimemagic/issues/97}, the \toolname{mimemagic} maintainer decides to resolve the license incompatibility by releasing new versions of \toolname{mimemagic} that switches license from \emph{MIT} to \emph{GPL-2.0}\citeurl{https://github.com/mimemagicrb/mimemagic/commit/c0f7b6b21a192629839db87612794d08f9ff7e88}, the same as the license of the required artifact.
Additionally, the maintainer decides to remove all previous releases with an incompatible license.
This change however, caused that the licenses of the web-framework \toolname{rails} and its dependency \toolname{mimemagic} are now incompatible with each other\citeurl{https://github.com/rails/rails/issues/41757}.
Essentially they were already before the license change, since \toolname{rails} depends on \toolname{xdg-shared-mime-info} indirectly, which would also require it to be under license compatible to \emph{GPL-2.0}.

The rails community discussed now how to resolve the license incompatibility issue on their side\citeurl{https://github.com/rails/rails/issues/41750}.
Since many commercial closed-source projects are built on top of \toolname{rails}, changing license to \emph{GPL-2.0} is not an option.
Developers discuss if they can replace the  dependency to \toolname{mimemagic} with another suitably licensed project.
Finally, after consulting the \toolname{mimemagic} developers, the latter decide to license their project again under \emph{MIT}, to remove the artifact that generated the issue from their sources, and to describe to developers how to obtain a copy of that artifact at runtime so that \toolname{mimemagic} can import it from the user's filesystem.

\textbf{Effects of License Incompatibilities:}
Imagine an e-government system that is built on top of \toolname{rails} by a third-party vendor and that \toolname{rails} were licensed under \gpl.
When the vendor hosts this system for a public sector institution, then it does not matter if any dependency is under \gpl, since it only prescribes open-sourcing a system on distribution.
That is, if the same vendor delivers the system to another party for hosting, the \gpl prescribes that the sources of the system are made publicly available.
Imagine now, that \toolname{rails} or its indirect dependency \toolname{mimemagic} were under \agpl.
That would mean that as soon as the system is publicly accessible, its sources would have to be made publicly available too, no matter which party is hosting the system.

That is the motivation for this work.
Usually, Danish e-government systems are closed source.
However, since they are most often web-applications and since they usually rely on third-party packages from various ecosystems, it is important to access how likely the license of a dependency would create an issue for online e-government systems.

\textbf{Terminology:}
Various ecosystems and package managers call reusable components differently.
For example, in the \maven realm reusable components are called \emph{Jars}, in \toolname{Ruby} they are calle \emph{Gems}, in \cargo they are called \emph{Crates} and in other ecosystems, e.g., \pypi and \npm they are just called \emph{packages}.
Irrespective the ecosystem, we call reusable components that are distributed via package managers uniformly \emph{packages}.

We call a dependency between two packages that are connected via a dependency relation in the dependency network a \emph{direct} dependency.
Contrary two packages possess an \emph{indirect} dependency if the required package is not directly declared by the dependent package but by an intermediate.

We call the act of making the sources of a software system publicly available under a corresponding license \emph{open-sourcing}.

Open-source software licenses fall into two major categories: \emph{permissive} and \emph{protective} licenses~\cite{kapitsaki2015insight}.
The latter are often called \emph{copyleft} licenses.

Permissive licenses restrict minimally how software can be used, modified, or redistributed.
Examples of such licenses are the \mymit, \license{Apache-2.0}, \license{BSD-3-Clause}, \license{ISC}, or the \license{Unlicense}.
Permissive licenses permit relicensing of derivative work and allow for use of software also in proprietary software that is distributed. 

There are two kinds of \emph{protective} licenses: \emph{weakly} and \emph{strongly} protective licenses.
Examples of weakly protective licenses are the \license{LGPL-3.0} or \license{MPL-2.0} and examples of strongly protective licenses are the \license{GPL-3.0} or \license{AGPL-3.0}. %
Strongly protective licenses prescribe to open source derived software and to distribute it under the same license\citeurl{https://www.gnu.org/licenses/copyleft.html}.
Software under a weak protective license can be redistributed under another license, as long as that software is not modified and software that reuses such does not need to be open-sourced\citeurl{https://www.gnu.org/licenses/license-compatibility.html}.

\section{Method}\label{sec:method}

\begin{figure}[b]
    \vspace{-1em}
    \centering
    \vspace{0.1em}
    \includegraphics[width=\linewidth]{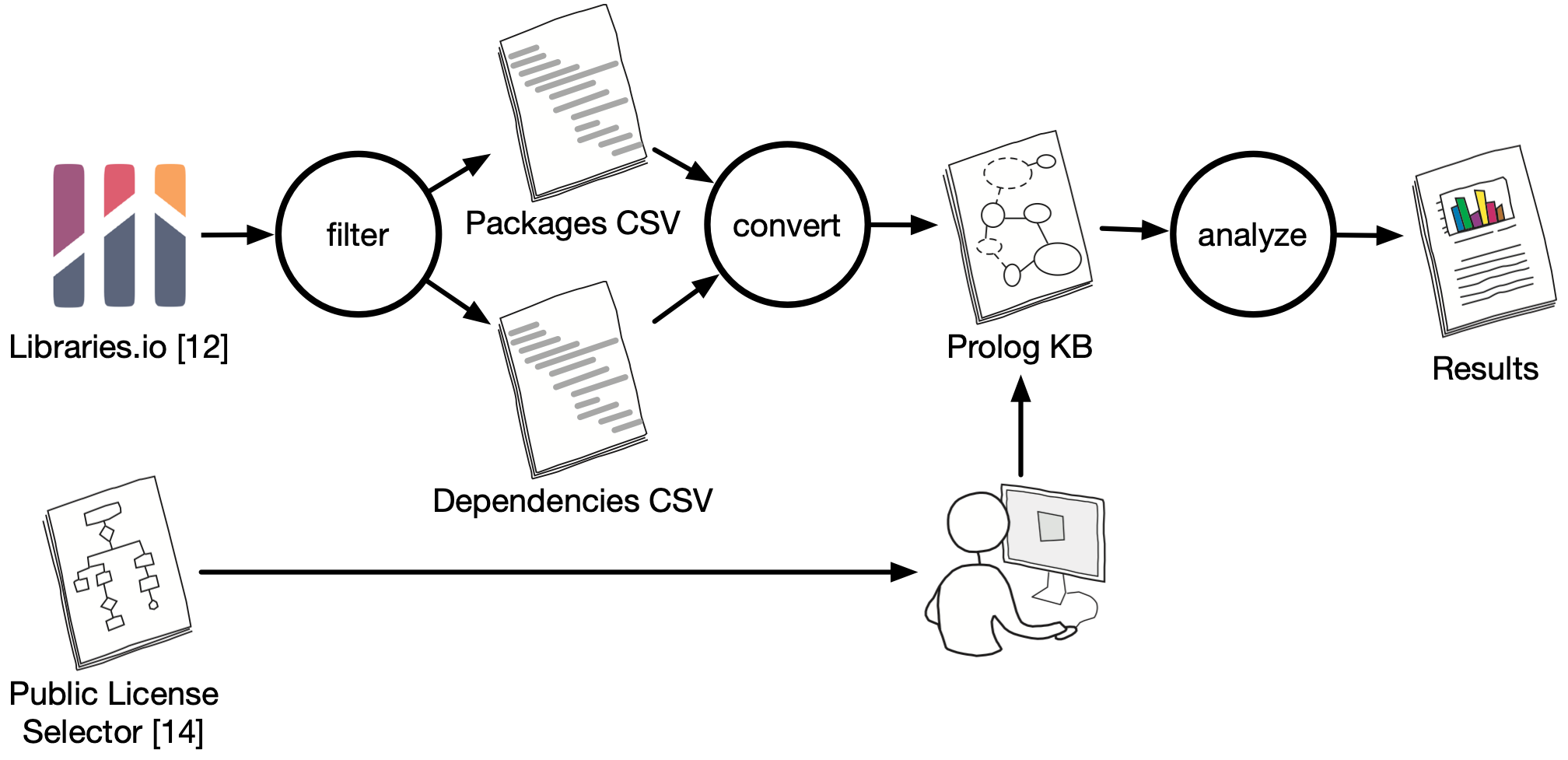}
    \vspace{-1em}
    \caption{Experiment setup.}
    \label{fig:setup}
\end{figure}

\autoref{fig:setup} illustrates the dataflow in our study setup.
A couple of \toolname{Bash}, \toolname{Python}, and \toolname{SWI Prolog} scripts automatically process and analyze the dependency networks of various software ecosystems and generate the results in the following section (\autoref{sec:results}).
A reproduction kit containing all these scripts and configurations is available online\citeurl{https://github.com/HelgeCPH/license_compat_study/}.

Our study is based on the \emph{Libraries.io} dataset~\cite{katz2020_3626071}. %
For various ecosystems, that dataset contains names, licenses, and dependency links of all registered packages.
Note, that the dataset provides only the license of the most \emph{current} release of a package, i.e., not the licenses of earlier versions of a package.
License information per package is stored via \gls{spdx} identifiers\citeurl{https://spdx.org/licenses/} --or a comma separated list of these--, which we use for this study.

We focus our study on the seven ecosystems \cargo, \maven, \npm, \nuget, \packagist, \pypi, and \ruby since \emph{a)} these ecosystems are most relevant for Danish e-government systems and \emph{b)} the \emph{Libraries.io} dataset contains dependency links for these ecosystems\footnote{It does not necessarily contain dependency links for all other ecosystems in the dataset}.

The \emph{filter} action in \autoref{fig:setup} illustrates that our scripts download the \emph{Libraries.io} dataset (more than 20GB in size) and extract all information relating to packages from the seven ecosystems
 and their dependencies.
Furthermore, the available data is reduced to retain only package identifiers, names, and license information.
We keep only those dependency links that are labeled as \emph{compile}- or \emph{runtime}-dependency in the original dataset since such dependencies may create license incompatibilities\footnote{Other dependencies types, e.g., \emph{test}, \emph{development} dependencies, etc. usually do not create a license incompatibility since they are not distributed with the resulting system.}.
Since the \emph{Libraries.io} dataset contains license information only on a package level, we reduce the dependencies between versions of packages to logical dependencies.
That is, we reduce the directed multigraph from the input dataset into a directed graph, e.g., as soon as one version of a package eventually depended on a version of another package, a single dependency link is retained.
The filtered data is stored for all seven ecosystems in two CSV files, one for the packages and another for the dependency links.

In a next step (illustrated as \emph{convert} action in \autoref{fig:setup}), these two CSV files are converted into \toolname{SWI Prolog} knowledge bases.
We do that, since license compatibility is a constraint satisfaction problem, which can be efficiently encoded and resolved with \toolname{Prolog} programs.
Manually, we create a set of license incompatibility rules as facts in a \toolname{Prolog} knowledge base.
\autoref{lst:prolog} shows \toolname{Prolog} facts that declare \license{AGPL-3.0} incompatible with the respectively stated license.
These \toolname{Prolog} facts are translated from the license incompatibility matrix from Kamocki et al.'s~\cite{kamocki2016public} \toolname{Public License Selector}\citeurl{https://github.com/ufal/public-license-selector/blob/97a7af0a7af00829bf43958669c79334cf77015c/src/definitions.coffee\#L257}.
During translation, we map the used license identifiers to the respective \gls{spdx} identifiers.

\begin{figure}[t]
\begin{lstlisting}[language=prolog, caption={Excerpt of \toolname{Prolog} license incompatibility rules.},label={lst:prolog}]
incompatible('AGPL-3.0', 'LGPL-2.1').
incompatible('AGPL-3.0', 'LGPL-2.1-only').
incompatible('AGPL-3.0', 'CDDL-1.0').
incompatible('AGPL-3.0', 'GPL-2.0').
incompatible('AGPL-3.0', 'GPL-2.0-only').
incompatible('AGPL-3.0', 'AGPL-1.0').
incompatible('AGPL-3.0', 'AGPL-1.0-only').
\end{lstlisting}
\end{figure}

\begingroup
\setlength{\tabcolsep}{2pt}
\begin{table*}[t]
\centering
\caption{Descriptive statistics of licenses and incompatibilities.}

\begin{tabular}{lrrrrrr}
\toprule
{}           & $\lvert \text{Packages} \rvert$ & $\lvert \text{Dependencies} \rvert$ & $\lvert \text{Packages}_{\text{discon}} \rvert$ & $\lvert \text{Packages}_{\text{con}} \rvert$ & $\lvert \text{Incompatibilities} \rvert$ & Incompatibilities\% \\
\midrule
Cargo     & \numprint{35635}   & \numprint{19968}   & \numprint{5918}   & \numprint{29717}  & \numprint{453}    & \numprint{2.3}\%  \\
Maven     & \numprint{184871}  & \numprint{426804}  & \numprint{101788} & \numprint{83083}  & \numprint{39002}  & \numprint{9.1}\%  \\
NPM       & \numprint{1275011} & \numprint{4927446} & \numprint{912541} & \numprint{362470} & \numprint{257593} & \numprint{5.2}\%  \\
NuGet     & \numprint{199447}  & \numprint{488385}  & \numprint{145996} & \numprint{53451}  & \numprint{22949}  & \numprint{4.7}\%  \\
Packagist & \numprint{313278}  & \numprint{473083}  & \numprint{162630} & \numprint{150648} & \numprint{43196}  & \numprint{9.1}\%  \\
PyPI      & \numprint{231690}  & \numprint{152779}  & \numprint{49197}  & \numprint{182493} & \numprint{31816}  & \numprint{20.8}\% \\
Rubygems  & \numprint{161608}  & \numprint{276580}  & \numprint{108552} & \numprint{53056}  & \numprint{17267}  & \numprint{6.2}\%  \\
\bottomrule
\end{tabular}
\label{tab:stats}
\end{table*}
\endgroup

\begin{figure}[b]
    \vspace{-1em}
    \centering
    \vspace{0.1em}
    \includegraphics[width=0.7\linewidth]{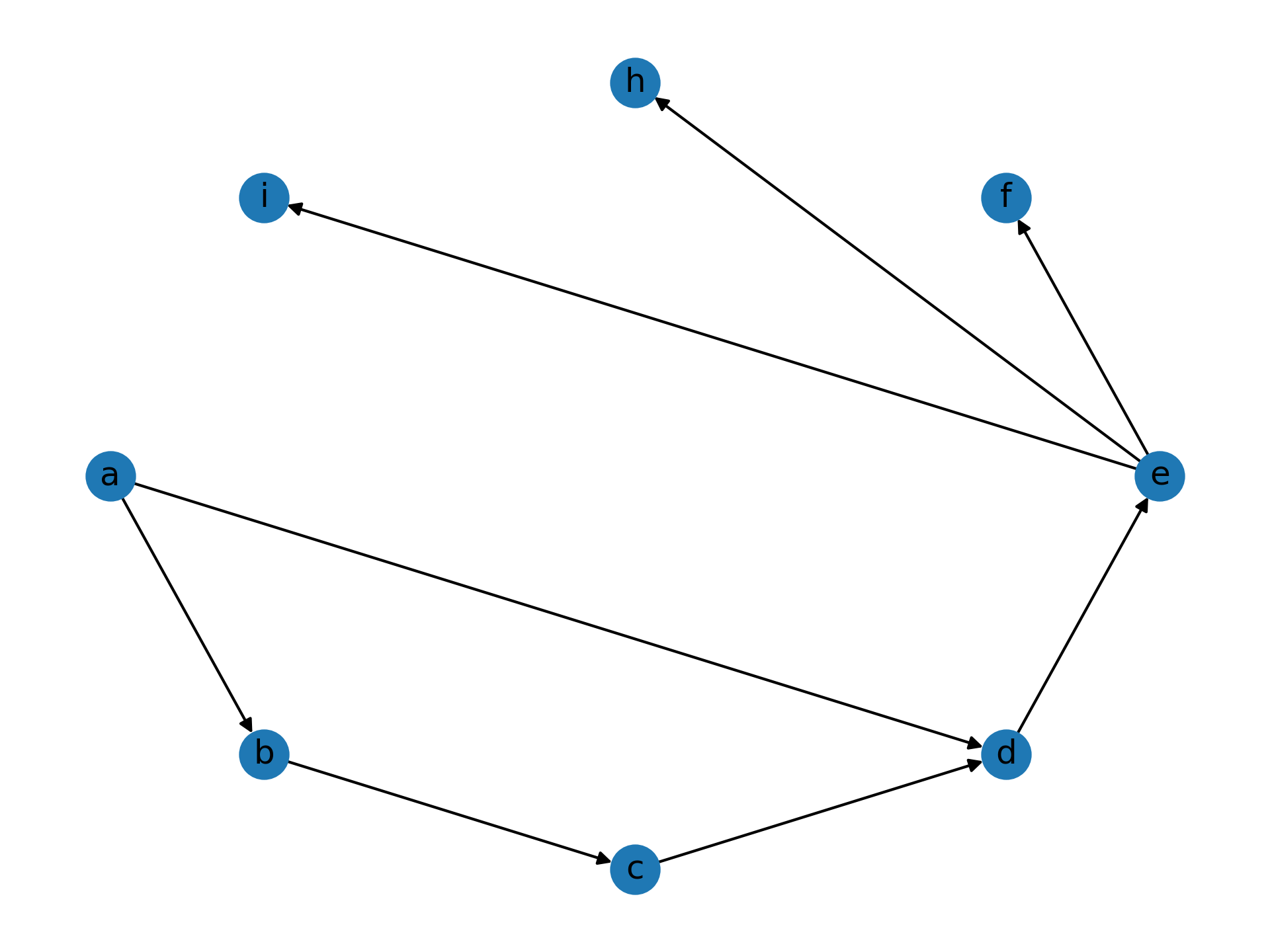}
    \vspace{-1em}
    \caption{Exemplary dependency network.}
    \label{fig:xmpl}
\end{figure}

To study RQ1 (\emph{Which licenses are mainly used in various software ecosystems?}), we just count frequencies of licenses per ecosystem and compute their relative frequency compared to the total number of packages in the relative ecosystem.
To investigate RQ2 (\emph{How many direct license incompatibilities exist in various ecosystems?}), we let the \toolname{Prolog} program compare all packages between which a dependency link exists.
For each such pair of nodes, the program checks if a license incompatibility rule --as in \autoref{lst:prolog}-- holds true.
For RQ3 (\emph{How many direct license incompatibilities are caused by \gls{agpl} licenses?}), we do the same as for RQ2 only that we restrict the analysis to dependency links that point to a package with any \gls{agpl} license, i.e., \license{APGL-1.0}, \license{APGL-2.0}, etc.

As mentioned above, packages in the \emph{Libraries.io} dataset can have more than one license assigned as a comma separated list of \gls{spdx} identifiers.
The semantics of a comma in a list of licenses is not specified in the dataset description\citeurl{https://libraries.io/data\#projectFields}.
We interpret such lists to prescribe a disjunction of possible licenses, i.e., a list of possible licenses to choose from depending on the use case.
Therefore, when searching for license incompatibilities, we analyze the cross-product of license combinations and record a license incompatibility only when all possible license pair combinations are incompatible.
Or vice versa, if one combination allows for a compatible use of a package, we do not mark the dependency as incompatible.

For investigation of RQ4 (\emph{What is the impact of indirect \gls{agpl} license incompatibilities?}), we want to incorporate the transitive structure of dependency relations.
Since license incompatibilities can propagate through a dependency network, as in the motivating example with \toolname{rails} and its indirect dependency \toolname{mimemagic} (\autoref{sec:background}), we identify all those packages that are under an \gls{agpl} and for which a dependent possesses an incompatible license.
Thereafter, we identify all packages that are on a path that ends in a package with a direct dependency to an \gls{agpl} licensed package with a license that is incompatible to an \gls{agpl} license.

Consider an exemplary dependency network as in \autoref{fig:xmpl}.
Let us assume that all packages (nodes) in that network are under \mymit license, except of packages \emph{i}, which is under \license{AGPL-3.0}.

For RQ1, we would report that $87.5\%$ of all licenses in the ecosystem are \mymit and that $12.5\%$ of all licenses are \license{AGPL-3.0}.

The \mymit license is incompatible with \license{AGPL-3.0}.
Consequently, there is a direct license incompatibility between package \emph{e} and package \emph{i}.
Therefore, we would report for RQ2 and RQ3 respectively, that there are \numprint{1} direct license incompatibility, i.e., $12.5\%$ of all dependency links point to a package where the dependency has a license that is incompatible with the dependent.

The impact of the incompatibility with \emph{i}'s \license{AGPL-3.0} is quite large since packages \emph{a}, \emph{b}, \emph{c}, and \emph{d} depend directly or indirectly on package \emph{e}.
That is, $50\%$ of the ecosystem depend on a package that reuses a package with a license that is incompatible to the respective own license.
By transitivity, they have an incompatible license too.

\begingroup
\setlength{\tabcolsep}{2pt}
\begin{table*}[t]
\centering
\caption{Descriptive statistics of licenses and incompatibilities.}

\begin{tabular}{lrrrr}
\toprule
{}        & $\lvert \text{Incompatibilites}_{\text{AGPL}} \rvert$ & $\text{Incompatibilities}_{\text{AGPL}}\text{\%}$ & $\lvert \text{Packages}_{\text{affected}} \rvert$ & $\text{Packages}_{\text{affected}}\text{\%}$ \\
\midrule
Cargo     & \numprint{0}   & \numprint{0.00}\% &     \numprint{0} & 0.00\%  \\
Maven     & \numprint{148} & \numprint{0.03}\% & \numprint{12236} & 6.62\% \\
NPM       & \numprint{777} & \numprint{0.02}\% & \numprint{30377} & 2.38\% \\
NuGet     & \numprint{26}  & \numprint{0.01}\% &    \numprint{52} & 0.03\% \\
Packagist & \numprint{66}  & \numprint{0.01}\% &   \numprint{120} & 0.04\% \\
PyPI      & \numprint{67}  & \numprint{0.04}\% &   \numprint{109} & 0.05\% \\
Rubygems  & \numprint{49}  & \numprint{0.02}\% &    \numprint{53} & 0.03\% \\
\bottomrule
\end{tabular}
\label{tab:stats2}
\end{table*}
\endgroup

\section{Results}\label{sec:results}

\begin{figure}[b]
  \centering
  \includegraphics[width=\linewidth]{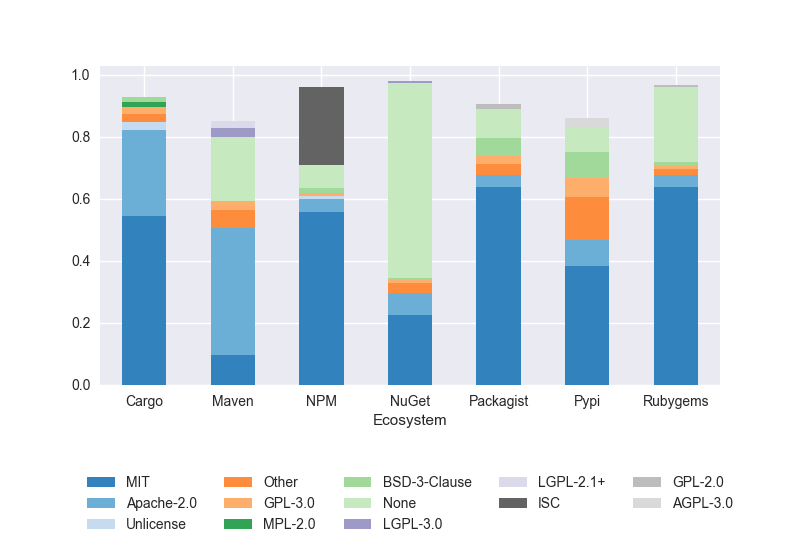}
\caption{Distribution of most common licenses per ecosystem.}\label{fig:licenses}
\end{figure}

\subsubsection{RQ1: Which licenses are mainly used in various software ecosystems?}\label{sec:rrq1}

\autoref{fig:licenses} illustrates the share of the most common licenses per ecosystem.
For clarity, we plot only the seven most common licenses per ecosystem.
A plethora of less frequent licenses --some of which are only used by a single package-- occupy the remaining percentages in \autoref{fig:licenses}. %

The \mymit license is most frequent in \cargo ($\approx54\%$), \npm ($\approx56\%$), \packagist ($\approx64\%$), \pypi ($\approx38\%$), and \ruby ($\approx64\%$).
For \maven, the \license{Apache-2.0} license is most common ($\approx41\%$) with a considerable share in \cargo ($\approx28\%$) too.
In the other five ecosystems the \license{Apache-2.0} license is less frequent with ca. $8\%$ in \pypi, ca. $7\%$ in \nuget, and ca. $4\%$ in \npm, \ruby, and \packagist respectively.
The only ecosystem in this study with a major share of \license{ISC} licensed packages is \npm where around a quarter of all packages carry it.

Generally, permissive licenses are most common in the ecosystems.
Except of \nuget, where the majority of packages do not carry any license, the majority of packages are permissively licensed.

A considerable amount of packages do not carry a license at all.
In \nuget the majority of packages ($\approx63\%$) do not carry a license.
The amount of such packages is lower in the other six ecosystems: \maven ($\approx20\%$), \ruby ($\approx24\%$), \pypi ($\approx14\%$), \packagist ($\approx9\%$), and \npm ($\approx7\%$).
In \autoref{fig:licenses} and in the \emph{Libraries.io} dataset packages without assigned license are labeled as \emph{``None''}.
Package without license are essentially not meant for reuse. %

Except of \pypi, strongly protective license are quite rare in the ecosystems.
\pypi sports the highest share of strongly protective licenses.
There, \license{GPL-3.0} and \license{AGPL-3.0} account together for $\approx9.2\%$ ($\approx6.1\%$ and $\approx3.1\%$ respectively).
In \packagist \license{GPL-2.0} and \license{GPL-3.0} are the common strongly protective licenses, together with a share of only $\approx4.2\%$.
For the remaining ecosystems, these numbers are even lower:
in \maven $\approx3.0\%$ of packages are \license{GPL-3.0} licensed,
in \cargo these are $\approx2.3\%$,
in \ruby packages with \license{GPL-2.0} or \license{GPL-3.0} together account for $\approx1.8\%$,
and \npm and \nuget have $\approx1.0\%$ of packages under \license{GPL-3.0} respectively.

Only in \pypi a noticeable share of packages is licensed under a version of \gls{agpl} (3.1\%, i.e., \numprint{7307} packages).
For the other ecosystems, this share is below one percent (\cargo: $0.78\%$, \maven: $0.74\%$, \npm: $0.24\%$, \nuget: $0.1\%$, \packagist $0.21\%$, and \ruby: $0.18\%$).

\begin{figure*}[t]
    \vspace{-1em}
    \centering
    \vspace{0.1em}
    \includegraphics[width=0.8\linewidth]{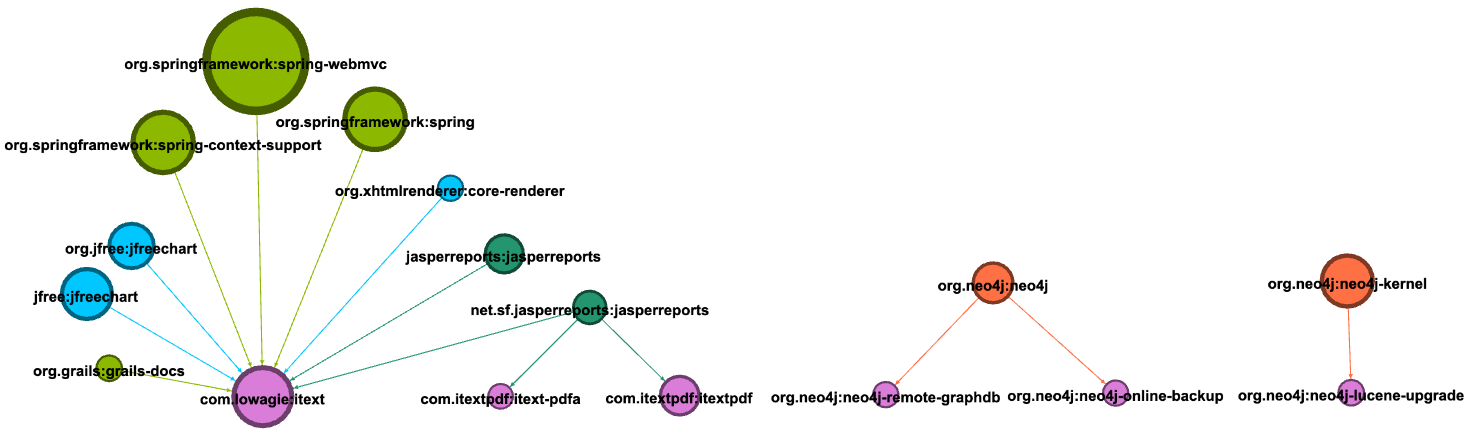}
    \vspace{-1em}
    \caption{Incompatibilities with \license{AGPL} licensed packages in \maven that pose major issues.}
    \label{fig:affected}
\end{figure*}

\subsubsection{RQ2: How many direct license incompatibilities exist in various ecosystems?}

\autoref{tab:stats} lists descriptive statistics for the seven studied ecosystems.
Column ``$\lvert \text{Packages} \rvert$'' shows the total number of packages, ``$\lvert \text{Dependencies} \rvert$'' shows the total number of logical dependency links in the dependency network (see \autoref{sec:background}), ``$\lvert \text{Packages}_{\text{discon}} \rvert$'' shows the total number of packages that neither depend on another package nor are required by any other package, ``$\lvert \text{Packages}_{\text{con}} \rvert$'' shows the number of packages that are connected as dependencies or dependents in the dependency network, and ``$\lvert \text{Incompatibilities} \rvert$'' and ``Incompatibilities\%'' show the absolute and relative amount of dependency links where the license of the dependency is incompatible with the license of the dependent package.

Obviously, \cargo is the smallest ecosystem with \numprint{35635} packages of which \numprint{29717} (ca. 83\% of all packages) contribute to the dependency network.
\npm is the largest ecosystem with \numprint{1275011} packages, where \numprint{912541} (ca. 72\% of all packages) are linked in the dependency network.
The sizes of the other five ecosystems are in between these two extremes, ranging from \numprint{161608} packages in \ruby to \numprint{313278} in \packagist.
The share of packages that are connected in the dependency network in the other ecosystems ranges from a meager 27\% in \nuget over 33\% (\ruby), 45\% (\maven), and 48\% \packagist to 79\% in \pypi.

\cargo sports the lowest amount of license incompatibilities with \numprint{2.3}\% of all dependencies that link packages with incompatible licenses (corresponding to \numprint{453} incompatibilities dependency links).
In \nuget these are \numprint{4.7}\% (\numprint{22949}).
In absolute numbers, \npm's \numprint{257593} license incompatibilities are the largest of all ecosystems.
But due to its total size, only \numprint{5.2}\% of all dependency links connect packages with incompatible licenses.
In the remaining ecosystems, the share of incompatible licenses raises from \numprint{6.2}\% in \ruby (\numprint{17267}) over \numprint{9.1}\% in both \maven (\numprint{39002}) and \packagist (\numprint{43196}) to a staggering \numprint{20.8}\% in \pypi (\numprint{31816}).

\subsubsection{RQ3: How many direct license incompatibilities are caused by \gls{agpl} licenses?}

\autoref{tab:stats2} lists the number of dependency links between packages where the dependency is under an \agpl license and the dependent is under an incompatible license (column $\lvert \text{Incompatibilites}_{\text{AGPL}} \rvert$ lists absolute numbers and $\text{Incompatibilities}_{\text{AGPL}}\text{\%}$ lists the ratio compared to all dependency links).

The only ecosystem without any license incompatibilities ``caused'' by dependencies to \agpl licensed packages is \cargo.
\npm and \maven have the highest absolute amount of such license incompatibilities with \numprint{777} and \numprint{148} respectively.
However, in general the ratio of incompatibilities caused by packages that incorrectly depend on \agpl licensed packages is low.
It is consistently below half per mill.
The highest relative share of incompatibilities caused by \agpl licensed packages have \pypi (0.04\%) and \maven (0.03\%).
Remember however, that \pypi is the ecosystem with the highest amount of \gls{agpl} licensed packages in this study, see \autoref{sec:rrq1}.

\subsubsection{RQ4: What is the impact of indirect \gls{agpl} license incompatibilities?}

Besides absolute and relative numbers of direct license incompatibilities that are caused by incorrectly depending on \agpl licensed packages, \autoref{tab:stats2} lists the number of packages that are affected by such an incompatibility (see columns ``$\lvert \text{Packages}_{\text{affected}} \rvert$'' for absolute and ``$\text{Packages}_{\text{affected}}\text{\%}$'' for relative numbers).
All packages that either directly or indirectly depend on a package with a license incompatibility are \emph{affected} by it.

The absolute and relative numbers of packages affected by an incompatibility with an \agpl license are non-existent in \cargo and very low for \nuget and \ruby (0.03\% with 52 and 53 affected packages respectively), for \packagist (0.04\%), and for \pypi (0.05\%).
For \npm, the share of packages that are affected by a license incompatibility with an \agpl licensed package is 2.38\%, which corresponds to \numprint{30377} packages in total.
However, these affected packages are fairly wide spread over the ecosystem.
In \npm none of the \agpl licensed dependencies with an incompatibly licensed dependent affects more than thousand other packages.
That is, it is not a few \agpl licensed packages that are dependencies of many other packages with an incompatible license.

To the contrary \maven is the ecosystem with the highest relative amount of \agpl-based license incompatibilities (6.62\%) with in total \numprint{12236} affected packages.
Here, a low number of packages that incorrectly depend on a package under \agpl license affects a large amount of other packages.
\autoref{fig:affected} illustrates all those packages from the \maven ecosystem (green, blue, and orange nodes on top) that depend on an \agpl licensed package (bottom nodes in pink) and that affect more than one thousand packages.
Sizes of nodes are adjusted to the \pr of the respective package in the \maven dependency network.
The larger a node the more central it is in the \maven ecosystem, i.e., the more packages depend either directly or indirectly on the respective package.
The three packages from the Spring framework\citeurl{https://spring.io/projects/spring-framework} (\toolname{org.springframework:spring-webmvc}, \toolname{org.springframework:spring-context-support}, \toolname{org.springframework:spring}) are all quite central to the dependency network.
They are respectively ranked to be the 171\textsuperscript{th}, 351\textsuperscript{st}, and 352\textsuperscript{nd} most central (important) nodes in the dependency network.
Less central but still fairly important are the \toolname{jfree}, \toolname{jasperreports}, \toolname{grails}, and \toolname{xhtmlrenderer} packages.
All of these depend on only three packages \toolname{com.lowagie.itext}, \toolname{com.itextpdf:itextpdf}, and \toolname{com.itextpdf-itext-pdfa}.
Each of these dependents affects more than \numprint{11500} other packages.

The two \toolname{neo4j} packages \toolname{org.neo4j:neo4j-kernel} and \toolname{org.neo4j:neo4j} are less central than the \toolname{springframework} packages in the dependency graph (ranks 465 and 761 respectively) but still, they affect more than \numprint{1300} packages each.
Interestingly, for these two packages the license incompatibility is caused by three packages (\toolname{org.neo4j:neo4j-remote-graphdb} and \toolname{org.neo4j:neo4j-lucene-upgrade}, \toolname{org.neo4j:neo4j-online-backup} respectively), which are created by the same organization.

Consequently, it is not too unlikely in \maven to indirectly depend on a package that has a license incompatibility with an \agpl licensed package.
Furthermore, in the \maven ecosystem we find a case similar to the example in the introduction, see \autoref{sec:intro}
Interestingly, like with \gls{rails} the Spring framework is also a web-development framework.

\section{Discussion}\label{sec:discussion}

As sketched in \autoref{sec:background}, there are multiple ways of resolving a license incompatibilities.
For example, a restrictive license of a dependency can be changed by the respective project owners to a more permissive license that is compatible to those of the dependents.
Alternatively, the dependent project changes its license from a more permissive to a more restrictive license that is compatible to the one of its dependency or it replaces the dependency all together with another suitably licensed package. 

Not all license incompatibilities are necessarily an issue.
For example, imagine a system that relies for operation on the monitoring tool \toolname{Prometheus}\citeurl{https://prometheus.io/}, which is licensed under the \license{Apache-2.0} license, and the dashboard visualization tool \toolname{Grafana}, which is licensed under the \license{AGPL-3.0} license.
Such system, does not have to be licensed under \license{AGPL-3.0} and be provided as open-source, since it is not derived work from \toolname{Grafana} but rather reuses a certain web-API during operation.
However, if this system was derived from \toolname{Grafana}, e.g., a modified version of it that is used to display data via a web-application, it would have to be open-sourced under a respective license.
These semantic differences of the purpose of package reuse cannot be inferred via static analysis, i.e., they are not represented in the results of this study.
The purpose of this study is only to access the size of a potential issue and to indicate potential refactoring candidates.

License incompatibilities as described in this paper may not be an issue in non-US American/Anglo-Saxon legislations.
For example, a French court ruled recently that \gpl copyright claims are invalid in France and may only be enforced via contractual disputes\citeurl{https://thehftguy.com/2021/08/30/french-appeal-court-affirms-decision-that-copyright-claims-on-gpl-are-invalid-must-be-enforced-via-contractual-dispute/}
Additionally, it is unclear if licenses, such as, the \gpl are actually a copyright or a contract in non US legislations~\cite{szattler2007gpl,guadamuz2004viral}.

\textit{Threats to validity:}

The quality of our results depend on the accuracy of dependency link information in the \emph{Libraries.io} dataset, which we trust to be accurate.

We are no lawyers.
That is, all reported results assume that the license incompatibility model, which we adapt from Kamocki et al.'s~\cite{kamocki2016public}, accurately captures incompatible licenses.
Likely this model is coarse grained, since license incompatibility also depends on the use case of software that reuses other packages.
Our model is use case agnostic and therefore might be too strict.

We rely on the latest available \emph{Libraries.io} dataset~\cite{katz2020_3626071}, which was released on Jan. 12th, 2020.
The dataset contains license information only statically on package level.
That is, for each package only one or more licenses are given.
However, a package might change its license(s) over time.
For example, versions of \toolname{com.lowagie:itext} below version 5.0.0 were released under the more permissive \license{MPL} or \license{LGPL}\citeurl{https://en.wikipedia.org/wiki/IText} and first with version 5.0.0 the license was switched to \license{AGPL-3.0}.
Since the most current, this is the only license that is recorded for the package.
Consequently, our results in \autoref{sec:results} may be overestimating \agpl/based license incompatibilities, since certain versions of packages that in our study exhibit license incompatibilities may depend on suitably licensed earlier versions.

The dependency links available in the \emph{Libraries.io} dataset are a lower bound approximation of the ``real'' dependency links in the ecosystems.
The dataset only includes those dependency links that are statically declared in the package metadata.
For example, \pypi and \maven packages can execute code during installation of packages, which can be used to install other dependencies on top of those that are statically declared.
These are not subject of this study.

The \emph{Libraries.io} dataset provides package licenses as comma separated lists of one or more licenses.
The semantics of such commas in a list of licenses is not specified in the dataset description\textsuperscript{16}.
Thus, we interpret that lists prescribe a disjunction of possible licenses, i.e., a list of possible licenses to choose from depending on the use case.
When searching for license incompatibilities, we analyze the cross-product of license combinations and record a license incompatibility only when all possible license pair combinations are incompatible.
That might be an overly permissive interpretation, since some projects want to express via multiple licenses that the provided sources are made up by source under the given licenses, i.e., a conjunction of these.

\section{Future Work}\label{sec:future}

In future, we plan to execute the experiment again on another dataset, which captures licenses per versions of packages and not per package in general.
This will allow us to assess if the reported high impact of license incompatibilities with a few \agpl licensed packages in \maven is an artifact or if it adequately represents reality.
We are currently working on creating such an updated dataset ourselves, since the \emph{Libraries.io} dataset appears to be stale\citeurl{https://github.com/librariesio/libraries.io/issues/2744}. 
Furthermore, working on more recent data will allow us to appropriately describe the current impact of \agpl licensed packages.

In our results, the amount of license incompatibilities that are caused by incorrect use of \agpl licensed packages is quite low or negligible for all ecosystems except \maven, see \autoref{sec:results}.
Likely, the reason for this is that the \agpl becomes increasingly popular first recently\citeurl{https://www.synopsys.com/blogs/software-security/using-agpl-adoption/}, i.e., after the latest release of the \emph{Libraries.io} dataset from Jan. 12th, 2020.

\section{Conclusions}\label{sec:conclusions}
In this paper, we investigate empirically and quantitatively to which degree license incompatibilities pose an issue in the seven major software ecosystems \cargo, \maven, \npm, \nuget, \packagist, \pypi, and \ruby.

Our results (\autoref{sec:results}) show that in most ecosystems the permissive \mymit and \apache licenses are most common, only in \nuget it is most common that packages do not carry any license.
\pypi is the only ecosystem with a noteworthy amount of packages under \license{AGPL-3.0} but only 3.1\% of all packages carry it.

The amount of license incompatibilities between packages that are directly connected via dependency links differs across the ecosystems.
The lowest amount of dependency links that cause a license incompatibility can be found in \cargo 2.3\% (453 dependencies).
This number increases from 4.7\% (\numprint{22949}) in \nuget over 5.2\% (\numprint{257593}), 6.2\% (\numprint{17267}), 9.1\% (\numprint{39002}), and 9.1\% (\numprint{43196}) in \npm, \ruby, \maven, and \packagist to 20.8\% (\numprint{31816}) in \pypi.
However, only a low amount of direct license incompatibilities are caused by packages that depend on \agpl licensed packages but possess an incompatible license themselves (max. 0.04\% in \pypi).

When studying the amount of packages that are either directly or indirectly affected by a license incompatibility with \agpl licensed packages, we find that 6.62\% (\numprint{12236}) of all \maven packages are affected but that only a few packages are causing such indirect license incompatibilities.

\subsection{Implications for Practitioners}

Our results illustrate that license incompatibilities in general and license incompatibilities caused by inappropriately depending on \agpl licensed packages are more frequent in some ecosystems compared to others.
Consequently, we recommend that practitioners not only check direct dependencies for compatible licenses but also indirect dependencies especially when working with \maven and \pypi.
Some package managers, e.g., \toolname{mvn} with the \toolname{dependency:tree} or \toolname{npm} with the \toolname{list} command, allow to inspect direct and indirect dependencies simultaneously.
In other package managers, such as, \toolname{pip}, dependency trees can be inspected via tools like \toolname{pipdeptree}\citeurl{https://github.com/naiquevin/pipdeptree}.
On top of that, there exist tools that can be integrated to the development process to support developers in creating software free of license incompatibilities, e.g., the IDE plugin \toolname{Sorrel}~\cite{pogrebnoy2021sorrel}.

\subsection{Implications for Researchers}

We only studied license incompatibilities within ecosystems.
Modern applications, in particular web-applications, typically reuse code from multiple ecosystems.
It remains an open question to which degree software projects from various domains are affected by license incompatibilities within ecosystems or to which degree they introduce even new incompatibilities on top of existing ones.

\bibliographystyle{IEEEtran}
\bibliography{bibliography}

\theendnotes

\end{document}